\documentstyle[12pt,aaspp4]{article} 
\newcommand  \kms   {km~s$^{-1}$} 
\newcommand  \vhel  {$V_{hel}$}
\newcommand  \nii   {[N~{\sc ii}]}
\newcommand  \ha    {H$\alpha$} 
\newcommand  \hii   {H\,{\sc ii}} 
\newcommand  \hi    {H\,{\sc i}}
\slugcomment{Appear to ApJ Vol. 503 (Aug 20)}

\begin{document}

\title{The Multi-Phase Medium in the Interstellar Complex N44}
\author{
Sungeun Kim,\altaffilmark{1}
You-Hua Chu,\altaffilmark{2}
Lister Staveley-Smith,\altaffilmark{3}
and
R. Chris Smith\altaffilmark{4}
}

\altaffiltext{1}{Mount Stromlo and Siding Spring Observatories, ANU, Weston 
Creek PO,
Canberra, ACT 2611, Australia}
\altaffiltext{2}{Astronomy Department, University of Illinois at 
Urbana-Champaign,
Urbana, IL 61801}
\altaffiltext{3}{Australia Telescope National Facility, CSIRO, P.O.\ Box
76, Epping, NSW 2121, Australia}
\altaffiltext{4}{Astronomy Department, University of Michigan, Ann Arbor, MI 
48109-1090}


\begin{abstract}

We have obtained high-resolution \hi\ observations of N44, one of 
the largest \hii\ complexes in the Large Magellanic Cloud.  The
distribution and internal motions of the \hi\ gas show dynamic 
effects of fast stellar winds and supernova blasts.  Numerous
\hi\ holes are detected, with the most prominent two corresponding
to the optically identified superbubbles Shell 1 and Shell 2.
The \hi\ gas associated with Shell 1 shows an expansion pattern
similar to that of the ionized gas shell, but the mass and kinetic 
energy of the \hi\ shell are 3--7 times those of the ionized gas
shell.  The total kinetic energy of the neutral and ionized gas 
of Shell 1 is still more than a factor of 5 lower than expected 
in a pressure-driven superbubble.  It is possible that the central
OB association was formed in a molecular cloud and a visible
superbubble was not fully developed until the ambient molecular gas
had been dissociated and cleared away.  This hypothesis is supported
by the existence of a molecular cloud toward N44 and the fact that
the apparent dynamic age of the superbubble Shell 1 is much shorter 
than the age of its OB association LH47.
Accelerated \hi\ gas is detected at the supernova remnant 0523$-$679.
The mass and kinetic energy in the associated \hi\ gas are also much 
higher than those in the ionized gas of 0523$-$679.  Studies of 
interstellar gas dynamics using ionized gas alone are clearly 
inadequate; neutral gas components {\em must} be included.

\end{abstract}

\keywords{ISM: bubbles - ISM: supernova remnant - Magellanic Clouds -
X-ray, HI, HII: ISM}

\section{Introduction}

N44, cataloged by \markcite{He56}Henize (1956), is a luminous \hii\
complex in the Large Magellanic Cloud (LMC).  It contains an
assortment of compact \hii\ regions, filaments, and shells of all
sizes, as well as three OB associations, LH47, 48, and 49
(\markcite{LH70}Lucke \& Hodge 1970).  As shown in Figure 1,  N44 is
dominated by a prominent shell around LH47 in the central region
(\markcite{CM90}Chu \& Mac Low 1990).  In
the surroundings are:  diffuse \hii\ regions and filaments to the
east;  compact \hii\ regions encompassing LH49 to the south; a large
faint shell to the west;  faint filaments to the north; a luminous,
compact \hii\ region around LH48 on the northeastern rim of the main
shell; and numerous small, bright, single-star \hii\ regions in the
outskirts of  N44.

Previous studies of ionized gas in N44 have yielded many interesting
results.  The main shell and the large shell to the west are designated 
as Shell 1 and Shell 2, respectively, and their expansion patterns have
been studied by \markcite{ML91}Meaburn \& Laspias (1991).   Shell 1
has been modeled as a superbubble using energy input implied  by the
observed massive stellar content; it is shown that the expansion
velocity of Shell 1 is much higher than expected  (\markcite{OM95}Oey
\& Massey 1995).   Diffuse X-ray emission has been detected in N44,
indicating the  existence of 10$^6$ K gas (\markcite{CM90}Chu \& Mac
Low 1990;  \markcite{WH91}Wang \& Helfand 1991). The X-ray emission
within  Shell 1 and Shell 2 has been suggested to be generated by
supernova remnants (SNRs) shocking the shell walls; the outward
extension of  X-ray emission at the southern periphery of Shell 1 has
been  interpreted as a breakout; and the bright diffuse X-ray source
at  6$'-7'$ northeast of Shell 1 has been identified as a new SNR
(\markcite{Ch93}Chu et al.\ 1993; \markcite{Ma96}Magnier et al.\ 1996).
Following the naming convention of using truncated B1950 coordinates,
this remnant will be named SNR 0523$-$679.

Interactions between massive stars and the interstellar medium 
(ISM) have led to the derivation of a multi-phase ISM model 
(\markcite{MO77}McKee \& Ostriker 1977) and have been used to 
explain the disk-halo connection in a galaxy (\markcite{NI89}Norman 
\& Ikeuchi 1989).  The physical structure of N44 vividly 
demonstrates the dynamical interaction between massive stars 
and the ISM.  Thus, N44 provides an excellent laboratory to study 
in detail the distribution, physical conditions, and relationship 
of the different phases of the ISM in a star forming region.

Previous studies of N44 have concentrated on only the warm and hot
ionized gas.  In order to carry out a comprehensive investigation 
of the multi-phase structure of N44, we have recently obtained 
high-resolution \hi\ observations of N44 with the Australia 
Telescope Compact Array (ATCA) and high-dispersion, long-slit 
echelle spectra of N44 with the 4m telescope at Cerro Tololo 
Inter-American Observatory.  In this paper we report these 
observations (\S 2), describe the interaction between \hi\ and 
\hii\ gases (\S 3), analyze the \hi\ gas associated with Shell 1 
and Shell 2 (\S 4), and study the acceleration of \hi\ gas 
by a SNR (\S 5).  A discussion is given at the end (\S 6).

\section{Observations}

\subsection{\hi\ Observations}

The \hi\ data of N44 are extracted from the \hi\ aperture synthesis 
survey of the LMC made with the Australia Telescope Compact Array 
(ATCA), which consists of 6 22-m antennas.
The details of this survey are given by Kim et al. (1997) and 
preliminary results are presented by Kim and Staveley-Smith (1997).   

In summary, the observations were made in four configurations:
750A on 1995 February 23 to March 11, 750B on 1996 January 27 to 
February 8, 750C on 1995 October 15 to 31, and 750D on 1994 October 26 
to November 9. These configurations each have five antennas with a 
maximum baseline of 750 m.  The combined configuration has 40 
independent baselines ranging from 30 to 750 m, with a baseline 
increment of 15.3 m.  The resultant angular resolution is $1\farcm0$ 
for the \hi\ images presented in this paper. The largest angular 
structure that the images are sensitive to is $\sim 0\fdg6$. 

The observing band was centered at 1.419 GHz with a bandwidth of 4 MHz.
The band was divided into 1024 channels; these channels were re-binned
into 400 channels after an on-line application of Hanning smoothing 
and edge rejection.  The final data cube used here has a velocity 
coverage of $190$ to 387 \kms, and a velocity resolution of 1.65 \kms.  
All \hi\ and H$\alpha$ velocities reported in this paper are heliocentric. 
 
The data cube of N44 was extracted from the full mosaic of the LMC, which 
consisted of 1344 pointing centers.  Approximately four pointing centers 
were included in the data cube of N44.  The phase and amplitude calibrators 
were PKS B0407$-$658 for some fields and PKS B0454$-$810 for the others. 
The primary ATCA calibrator, PKS B1934$-$638 (assumed flux density 14.9 Jy 
at 1.419 GHz), was observed at the start and end of each observing day. 
These observations served both for bandpass calibration and flux density 
calibration. 

The data were edited, calibrated, and mosaicked in {\sc miriad}. The pixel 
size is 20$''$ and the size of the N44 cube is $23\times23\times120$.  
Superuniform weighting (Sramek and Schwab 1989) was applied to the $uv$ data 
with an additional Gaussian taper.  The resulting data were then Fourier 
transformed and mosaicked in the image plane.   The maximum entropy method 
was used to deconvolve this cube. The final cube was constructed by convolving
the maximum entropy model with a Gaussian of FWHM 1$'\times1'$. The final 
rms noise, measured in line-free parts of the final cube, is 9K.


\subsection{Echelle Observations}

To examine the kinematic properties of the ionized gas in N44,  we
obtained high-dispersion spectroscopic observations in the  \ha+\nii\
lines with the echelle spectrograph on the 4m telescope  at Cerro
Tololo Inter-American Observatory on 1995 January 18.  Using a
post-slit \ha\ filter and replacing the cross-disperser with a flat
mirror, we were able to observe both \ha\ and
\nii$\lambda\lambda$6548, 6583 lines over a long slit.   The 79 lines
mm$^{-1}$ echelle grating and the red long focus camera were used.
The data were recorded with a Tek  2048$\times$2048 CCD.  The 24
$\mu$m pixel size corresponds to about 0.82 \AA\ (3.75 \kms\ at the
\ha\ line) along the spectral  direction and 0\farcs267 along the
spatial direction.  The  spectral coverage was limited by the filter
width to 125 \AA.   The spatial coverage was limited by the slit
length to $\sim4'$,  with some vignetting in the outer
1\farcm5.  The slit width was 250 $\mu$m, or 1\farcs66, and the
instrumental  FWHM was 16.1$\pm$0.8 \kms.

The wavelength calibration and geometric distortion correction were
carried out with the use of Th-Ar lamp exposures in the beginning of
the night.  The observations were sufficiently deep that the
geocoronal \ha\ component was detected in each frame.  The geocoronal
\ha\ component (at zero observed velocity) provided an accurate and
convenient reference for velocity measurements.  The observed
velocities were then converted to heliocentric velocities.

The echellograms are shown in Figure 1.  Only \ha\ line is presented
in the echellograms.  The \nii\ lines are detected at a much lower S/N
ratio than the \ha\ line, hence they are not presented here.

\subsection{Optical Imaging}

In order to compare the neutral gas components to the ionized gas in
the N44 region, we obtained optical H$\alpha$ emission-line images
with a CCD camera mounted at the Newtonian focus of the Curtis Schmidt
telescope at CTIO on 18 February 1994 UT. The detector was a
front-illuminated Thomson $1024\times1028$ CCD with 19$\mu$ pixels,
giving a scale of 1\farcs835 pixel$^{-1}$ and a field of view of 
31\farcm3.  A narrow-bandpass H$\alpha$ filter ($\lambda_c = 6561$ 
\AA, $\Delta\lambda = 26$ \AA) was used to isolate the emission, and
a red continuum filter ($\lambda_c = 6840$ \AA, $\Delta\lambda = 95$
\AA) was used to obtain images of the continuum background (images in
other emission lines were also obtained, but those will be discussed
in another paper).  Multiple frames were obtained through each filter,
amounting to total integration times of 1200s in H$\alpha$ and 600s
in the continuum.  The data were reduced with IRAF\footnote{IRAF is
distributed by the National Optical Astronomy Observatories (NOAO).},
and multiple frames were shifted and combined to obtain the images
shown in Figures 1 and 3b.

\section{Interaction between \hi\ and \hii\ Components}

The integrated \hi\ map of N44 (Figure 2) shows two clear
depressions that correspond to the interiors of the optically
defined Shell 1 and Shell 2, and peaks that are adjacent to
\ha\ features.  The relationship between the \hi\ gas and the 
\hii\ gas is clearly complex.  The ATCA channel maps of \hi\ 
(Figure 3a) kinematically resolve the distribution of \hi\ along 
the line of sight, and allow more precise identification 
of physical structures in the \hi\ distribution.  Comparing
these iso-velocity maps of \hi\ to the \ha\ image (Figure 3b), 
we may examine the relationship between the \hi\ and \hii\ gas 
in greater detail and accuracy.  

 From the comparisons of \hi\ and \ha\ images, we have found
indications of: (1) \hi\ holes and expanding shells, (2)
acceleration of \hi, (3) compression of \hi, and (4) 
the opening of a breakout in \hi. These phenomena are described 
below in more detail.

{\bf \hi\ holes and expanding shells.}
The H$\alpha$ image in Figure 1 reveals the prominent superbubble
Shell 1 and the fainter superbubble Shell 2 in N44.  Shell 1, around
the OB association LH47, has a size of 70 pc $\times$ 50 pc; while
Shell 2, with no obvious OB association inside, is 60 pc across.
The \hi\ channel maps at \vhel\ = 292 -- 302 \kms\ (Figure 3) show 
clear \hi\ holes at the central cavities of Shell 1 and Shell 2.
Projected within the \hi\ holes are high-velocity \hi\ clumps
($\sim$15 pc diameter) at 5$^{\rm h}$22$^{\rm m}$45$^{\rm s}$,
$-67^\circ56'00''$ (J2000) at the velocity range of \vhel\ 
$\sim$ 228 -- 245 \kms (see Figure 3). These high-velocity \hi\ clumps 
are most likely the brightest parts of the approaching hemisphere of an 
expanding \hi\ shell associated with the optical shell.  The receding 
hemispheres of these shells expand more slowly.  The expansion of these 
\hi\ shells is better visualized in the position--velocity plots (or $L-V$ 
diagrams) in Figure 4.  These shells will be analyzed in detail in \S 4.  

Besides the \hi\ shells with obvious optical counterparts,
we see two \hi\ holes, centered at \vhel\ = 297 \kms, without known 
expanding shells in the optical.  Inspected closely, the \hi\ hole at 
5$^{\rm h}$22$^{\rm m}$00$^{\rm s}$, $-67^\circ51'50''$ (J2000) 
is associated with sharp, curved optical filaments that appear to
delineate a shell structure, while the \hi\ hole at 
5$^{\rm h}$21$^{\rm m}$07$^{\rm s}$, $-67^\circ51'45''$ (J2000)
appears to be associated with only faint, irregular optical filaments.
Interestingly, as shown in Figure 5, both of these \hi\ holes 
encompass regions of bright diffuse X-ray emission reported by Chu 
et al.\ (1993).  The relationship between the X-ray-emitting ionized 
gas and the \hi\ gas will be discussed further in \S 6.

{\bf Acceleration of \hi.}
The \hi\ channel maps in Figure 3 show that the bulk of \hi\ gas
is detected in the velocity range of 295 -- 305 \kms.
High-velocity \hi\ gas with $\Delta V$ up to $-$80 \kms\ is 
detected in a portion of the SNR 0523$-$679.
This high-velocity gas must have been accelerated by the SNR.  
High-velocity \hi\ gas (\vhel\ = 233 \kms) is also detected in the 
region between the SNR and the superbubble Shell 1.  This gas might 
consist of two components, one accelerated by the SNR and the other
by Shell 1, as the \hi\ contours have an excellent correspondence
with an \ha\ blister on the periphery of Shell 1 .

{\bf Compression of \hi.}
In the channel maps near the bulk velocity, \hi\ peaks are seen in 
regions between expanding \hi\ shells.  It is possible that these 
peaks represent compressions caused by two shells expanding into 
each other.  One example is the region at 
5$^{\rm h}$22$^{\rm m}$40$^{\rm s}$, $-67^\circ54'40''$ (J2000) 
between the SNR 0523$-$679 and the superbubble Shell 1; the \hi\ 
column density, $N$(\hi) = 3.9 $\times$ 10$^{21}$ cm$^{-2}$, is
a factor of 2 higher than those of adjacent regions along the 
rim of shell.\footnote{The \hi\ velocity profile of this region has an 
extended red wing up to \vhel\ $\sim$ 320 \kms.  However, the channel
map centered at 310 \kms\ shows an unrelated feature extending from
this region to the northeast. The compression is referred to the
dense clump at the systemic bulk velocity.}
Another example is the region at 5$^{\rm h}$23$^{\rm m}$06$^{\rm s}$,
5$^{\rm h}$21$^{\rm m}$58$^{\rm s}$, $-67^\circ55'$00$''$ (J2000)
bounded by Shell 1, Shell 2, and the \hi\ hole to the north.
The \hi\ column density, $N$(\hi) = 3.5 $\times$ 10$^{21}$ cm$^{-2}$, 
is a factor of 2 higher than those of adjacent regions along the rim of
shell. It is unlikely that the \hi\ peaks are caused by a low ionization 
in these directions, since the ionizing source of Shell 1 is located at 
the center of the shell.  These \hi\ peaks must be real enhanced column 
densities as a result of compression between expanding shells. 

{\bf Opening of a Breakout in \hi.}
A breakout in Shell 1 toward the south has been suggested by Chu et al.\
(1993), based on an extension of diffuse X-ray emission toward the 
group of \hii\ regions associated with the OB association LH49.
This explanation has been confirmed by the presence of high-velocity 
ionized gas and a lower plasma temperature in the breakout region 
(Magnier et al.\ 1996).  The \hi\ channel maps at \vhel\ of 300 -- 305
\kms\ show a clear extension of the \hi\ hole to the southeast, 
coinciding with the X-ray extension.  This extension of the \hi\ hole 
is likely the opening through which the breakout occurs.
Near the southeast end of the X-ray extension, we see an \hi\ clump
peaking at \vhel\ = 285 \kms\ with a column density $N$(\hi) of 
2.8 $\times$10$^{21}$ cm$^{-2}$, a factor of 3 higher than the 
surrounding region.  This \hi\ clump might have been impinged upon
and compressed by the flow of hot, ionized gas in the breakout.

\section{The Superbubble Structure of Shell 1 and Shell 2}

Previous studies of the superbubble structure of N44 have used 
observations of only ionized gas (e.g., Chu et al.\ 1993; Magnier
et al.\ 1996).  It is found that the observed kinetic energy of
the ionized Shell 1 of N44 is at least an order of magnitude lower
than the value expected in Weaver et al.'s (1977) superbubble 
model (Magnier et al.\ 1996).  Our new \hi\ observation of N44
reveals the existence of a neutral gas associated with Shell 1
and possibly Shell 2 as well.  The kinetic energy of the neutral
shell may alleviate the aforementioned discrepancy in energies.
Below we derive the dynamic parameters separately for the 
ionized and neutral components of Shell 1.  The energetics of 
Shell 2 is discussed near the end of this section.

The expansion of Shell 1 is not uniform.  Its systemic velocity, 
determined from the \ha\ velocities at the rim of the 
shell, is 295 -- 298 \kms.  Referenced to this systemic velocity, 
the receding hemisphere of the ionized gas shell has an expansion 
velocity of 30 \kms, and the approaching side 45 \kms.
We will analyze the physical parameters of the approaching and 
receding hemispheres separately.  These parameters are summarized 
in Table 2. 

The \ha\ echelle observation of the central region of Shell 1 
shows that the receding component is three times as bright as the 
approaching component.  Using photoelectrically calibrated PDS 
scans of the Curtis Schmidt plates of \markcite{KH86}Kennicutt 
\& Hodge (1986) and together with our CCD \ha\ image, we
derive an emission measure of 214 cm$^{-6}$ pc for the receding
component and 71 cm$^{-6}$ pc for the approaching component.
To derive the rms shell density, we have assumed a uniform thin shell
for Shell 1 to estimate its shell thickness.  The \ha\ surface 
brightness profile of Shell 1 shows a peak-to-center ratio of 13,
indicating a fractional shell thickness $\Delta R / R$ of 0.02.
The FWHM of the brightness peak at the shell rim is 0.1 times the
shell radius.  Given the nonuniform densities, the fractional shell
thickness is most likely within the range of 0.02 -- 0.1.  We have 
derived the rms density, mass, and kinetic energy of the shell for
shell thicknesses of 0.02 and 0.1, respectively, and listed them
in Table 2.  
The rms density in the receding hemisphere is almost twice as high 
as that in the approaching hemisphere, but the kinetic energy of 
the receding hemisphere is 21\% lower than that of the approaching
hemisphere.
The total mass and kinetic energy of the \hii\ shell 
are 2580 $M_\odot$ and 3.5 $\times$ 10$^{49}$ ergs for a 
$\Delta R / R$ of 0.1, or 1210 $M_\odot$ and 1.6 $\times$ 10$^{49}$ 
ergs for a $\Delta R / R$ of 0.02.

The \hi\ material associated with Shell 1 of N44 shows a typical 
expansion structure in the position--velocity diagrams (Figure 4) 
along cuts crossing the central portion of Shell 1.  \hi\ is 
detected in the velocity range of \vhel\ $\sim$ 220 -- 327 \kms. 
The velocity gradient of the high-velocity component in the central 
cuts of Shell 1 indicates an expansion motion.  At the center of 
Shell 1, the receding side of the \hi\ shell is expanding at a 
lower velocity, 28$\pm$5 \kms, than the approaching side, 50$\pm$5 
\kms.  These expansion velocities are comparable to those of the 
ionized component of Shell 1.

Following the analysis of the ionized shell, we discuss the receding
and approaching hemispheres of the \hi\ shell separately.  The column
densities, masses, and kinetic energies are listed in Table 2.
The \hi\ column density of the receding hemisphere is 2.5 times that
of the approaching hemisphere, but the kinetic energy of the receding
hemisphere is 21\% lower.  This contrast is remarkably similar to that
in the \hii\ shell.  The total mass of the \hi\ shell is 8640 $M_\odot$,
and the total kinetic energy in \hi\ is 1.1 $\times$ 10$^{50}$ ergs.

Note that the total mass and total kinetic energy of the \hi\ component
of Shell 1 are 3 -- 7 times higher than those of the \hii\ component.  This
implies that analyses of superbubble dynamics {\it must} include the
neutral gas component!  Note also that the asymmetric expansion of
Shell 1, shown in both \hi\ and \hii\ components, indicates a stratified
interstellar medium with densities decreasing toward us.

We may re-examine the energetics of Shell 1 with the additional
\hi\ information.  The total stellar wind and supernova energy input
implied by the observed stellar content (Oey \& Massey 1995), the total 
kinetic energy derived in this paper, and the thermal energy of the hot 
interior derived from X-ray observations (Magnier et al.\ 1996) are
summarized in Table 2.  In a pressure-driven superbubble, the thermal 
and kinetic energies are expected to be 35/77 and 15/77 times the
total mechanic energy input from stellar winds and supernovae
(Weaver et al. 1977; \markcite{MM88}Mac Low \& McCray 1988).
The expected thermal energy and shell kinetic energy are also listed 
in Table 2.  The observed shell kinetic energy is dominated by that 
of the expanding \hi\ shell; still, there is at least a factor of 5
discrepancy between the observed and expected kinetic energies.  
The discrepancy might be caused by the breakout of Shell 1, since 
a large fraction of the stellar energy input may have been lost in 
the breakout.  Other possibilities are discussed in \S 6.2.

The superbubble Shell 2 of N44 appears as an \hi\ hole in the channel 
maps as well as in the position-velocity diagrams (Figure 4).  The 
receding side of the shell is detected along some cuts across the 
shell.  The approaching side is much fainter; only a bright, 
high-velocity \hi\ knot is detected.  (The rest of the approaching 
hemisphere is probably below our detection limit.)

In a similar way, we describe the physical parameters of the \hi\
component of Shell 2.  The receding part of the \hi\ shell has an 
expansion velocity of 30$\pm$5 \kms, lower than that of the 
approaching side, 50$\pm$5 \kms.
The asymmetric expansion of the \hi\ in Shell 2 is similar to that of
Shell 1; therefore, Shell 1 and Shell 2 probably share the same 
interstellar environment.
The \hi\ mass of Shell 2 is 5850 $M_\odot$ in the receding hemisphere
and 3090 $M_\odot$ in the approaching hemisphere.  The total 
kinetic energy of the \hi\ in Shell 2 is 1.3 $\times$10$^{50}$ ergs,
20\% higher than that of the \hi\ in Shell 1.
The stellar content of Shell 2 has not been as well studied as that
of Shell 1, and the X-ray observations of Shell 2 have much lower
S/N ratios, hence Shell 2 cannot be modeled in as much detail as Shell 1.

\section{The Physical Structure of the SNR 0523$-$679}

The SNR 0523$-$679, centered at 
5$^{\rm h}$23$^{\rm m}$06$^{\rm s}$, $-67^\circ53'15''$ (J2000),
was first diagnosed by its diffuse X-ray emission (Chu et 
al.\ 1993).  The diffuse X-ray emission is surrounded by curved
\ha\ filaments that are suggestive of a shell with a diameter of 
$\sim$3\farcm8.  The \ha\ shell, called Shell 3 by Chu et al.\ (1993),
is bounded by \hii\ regions on the west and south sides.  The 
bright arc on the west side of the shell
is also on the surface of the \hii\ region around the OB association LH48,
indicating a physical interaction between the SNR and the \hii\ region.
The southern part of the SNR shell is blended with a diffuse ring-like
\hii\ region; however, it is not clear whether they physically interact.
Within the boundary of the SNR shell, long \ha\ filaments and dust lanes 
run from the NE corner to the SW corner, dissecting the shell into two lobes.
The X-ray emission shows a corresponding 2-lobe structure, which 
could be caused by the absorption in the dust lane.

The internal motion of the SNR is revealed in the echellograms at slit 
positions S4 and S5 (Figure 1).  The velocities at the shell rim sampled 
by these two slits are \vhel\ = 297 -- 304 \kms, which will be adopted as 
the systemic velocity of the ionized SNR shell.  The echellograms show
very different velocity structures in the north lobe and the south lobe 
of the SNR.  The north lobe shows prominent, blue-shifted material in 
the \ha\ line; the brighter parts have velocity offsets up to $-$125 \kms, 
while the fainter parts have velocity offsets up to $-$150 \kms.
The receding side of the north lobe has much smaller velocity offsets,
$\sim$20 \kms.
The south lobe shows mostly red-shifted material in the \ha\ line, with
velocity offsets up to +100 \kms; however, blue-shifted material is also
detected at the northern end of the south lobe, with velocity offsets up
to $-$100 \kms.

For such an irregular pattern of expansion, it is impossible to 
determine the kinetic energy of the ionized gas in SNR 0523$-$679 
using our limited echelle observations.  We can nevertheless
assume a uniform shell and derive upper limits on the mass and 
kinetic energy.  The brightest part of the SNR along slit S5 has
an emission measure of 25 cm$^{-6}$ pc.  The apparent width of the
SNR shell, 6\farcs5 or 1.6 pc, provides an upper limit on the real
shell thickness.  The corresponding rms $n_e$ is $\sim$ 4 cm$^{-3}$.
The upper limits on the mass and kinetic energy are thus 370 M$_\odot$
and 6 $\times$ 10$^{49}$ ergs.

The \hi\ gas associated with SNR 0523$-$679 has a very different 
velocity structure from that of the ionized gas described above.  
The bulk velocity of \hi\ gas toward the SNR is \vhel\ = 306$\pm$5 
\kms.  Since this velocity is similar to the SNR's systemic velocity 
determined with the \ha\ line, we will adopt this velocity as the 
systemic velocity of the \hi\ gas associated with the SNR.  The 
channel maps in Figure 3 show an \hi\ hole at \vhel\ = 297 \kms, 
although the size of the \hi\ hole is smaller than that of the \ha\ 
shell.

The receding side of the \hi\ gas associated with the SNR is clearly 
seen in the channel map at \vhel\ = 325 \kms, suggesting an expansion 
velocity of 20--30 \kms.   The approaching side of the \hi\ gas is more 
complex.  Blue-shifted \hi\ gas in the velocity range \vhel\ = 
230 -- 245 \kms\ is seen toward the SNR.  However, this \hi\ gas also 
extends further south than the boundary of the SNR; the southern 
extension is particularly pronounced in the region between the SNR and 
Shell 1 at \vhel\ = 228 -- 235 \kms.  As we explained in \S3, Shell 1 
might be responsible for the acceleration of the high-velocity \hi\ in 
the region between Shell 1 and the SNR. Interestingly, the 
highest-velocity (\vhel\ = 226 \kms) \hi\ component unambiguously 
associated with the SNR is located at the NE edge, instead of the 
center, of the optical shell.  This might indicate that a breakout 
similar to that seen at the south edge of Shell 1 has occurred at 
this position.

We may estimate the mass and kinetic energy of the \hi\ gas associated
with the SNR.  The high-velocity \hi\ gas is more extended than the 
optical boundary of the SNR. To exclude the contribution from other 
sources, we use only the part of \hi\ gas projected within the optical 
boundary of the SNR. The mass of the approaching side of the \hi\ 
gas is 7460$\pm$1000 $M_\odot$ and the kinetic energy of the
SNR is 2.7 $\times$ 10$^{50}$ ergs.   The mass and kinetic energy of 
the \hi\ gas are much higher than those of the ionized gas in the SNR.
This is similar  to what we have found for the superbubble Shell 1.

The thermal energy of the SNR can be derived from the ROSAT and ASCA
observations in X-rays.  The best Raymond and Smith (1977) model fits 
of these two sets of data give very different plasma temperature and
absorption column density; however, the thermal energies derived from
these two sets of parameters are similar.  The best fit to the ROSAT
Position Sensitive Proportional Counter data (Chu et al. 1993) gives
a plasma temperature of $kT$ = 0.835 keV, an absorption column
density of log $N_H$ = 20.68, and a normalization factor of log
$N_e^2 Vf/4\pi D^2$ = 10.817, in the cgs units, where $N_e$ is the electron
density, $V$ is the volume, $f$ is the volume filling factor, and $D$ is the
distance. The thermal energy of the SNR calculated from this model
fit is 1.2 $\times$ 10$^{50}$$(f/0.5)^{1/2}$ ergs. The best fit to the
ASCA Solid-state Imaging Spectrometers (SIS) data gives $kT$ = 0.35
keV, log $N_H$ = 21.48, and log $N_e^2 Vf/4\pi D^2$ = 11.326. The
thermal energy from this model fit is 9.0 $\times$
10$^{49}$$(f/0.5)^{1/2}$ ergs.  Considering the uncertainties of
filling factor and spectral fits, the thermal energy of the SNR is
probably (1.0$\pm$0.5)$\times$ 10$^{50}$ ergs. 

The thermal energy in SNR 0523$-$679 is similar to that in each of 
the shells in the SNR DEM\,L\,316 (\markcite{Wi97}Williams et al. 
1997), despite the fact that SNR 0523$-$679 is in a much more 
inhomogeneous medium and has a much more nonuniform expansion.  
The kinetic energy in the ionized gas of SNR 0523$-$679 is of a 
similar order of magnitude as the thermal energy of SNR 0523$-$679.  
This behavior is similar to what is observed in DEM\,L\,316.
The total kinetic energy (\hi\ + ionized) of SNR 0523$-$679 is much
larger than the thermal energy in the SNR's hot interior.  This 
behavior is similar to what is seen in N44's superbubble Shell 1.

\section{Discussion}

\subsection{Implications on the \hi\ Structure of the LMC}

The ATCA \hi\ survey of the LMC (Kim \& Staveley-Smith 1997) shows
\hi\ holes with sizes ranging from a few tens of pc to $\sim$1400 pc.
It is conceivable that these \hi\ holes have been created by the
combined effects of fast stellar winds and supernova explosions.  
However, most of these holes do not have any recognizable stellar 
content or corresponding ionized gas shells, making follow-up studies 
difficult, if not impossible.  N44, containing three OB associations 
at three different evolutionary stages, offers an ideal site for 
us to study how massive stars shape the interstellar medium.  The 
current \hi\ structure in N44 is readily visible, whereas the ionized 
gas structure allows us to foresee the future \hi\ structure after
the ionizing fluxes have terminated at the demise of all massive
stars in N44.  From the study of N44, we hope to gain insight into 
the general \hi\ structure of the LMC.

We first examine the validity of the apparent ``holes" in the \hi\ 
maps of N44.  The integrated \hi\ map of N44 shows two deep 
depressions.  Both depressions have associated \hi\ shells, and 
both shells have optical counterparts, the superbubbles Shell 1 
and Shell 2.  For Shell 1, the \hi\ shell
and the ionized gas shell have similar expansion patterns and 
velocities: both have the far side receding at 30 \kms\ and the 
near side approaching at 45--50 \kms.  For Shell 2, the \hi\ shell 
appears to expand faster than the ionized gas shell: the \hi\ shell 
has expansion velocities of 30--50 \kms, while the ionized gas shell
show only a velocity variation from 304 to 320 \kms\ (Meaburn \& 
Laspias 1991).  This different behavior might be caused by the
different distribution of ionizing sources.  Shell 1 has a central
ionizing source, the OB association LH47, hence its neutral \hi\ shell
is adjacent and exterior to its ionized gas shell.  Shell 2, on the
other hand, has no central OB association, and its ionizing flux might
be provided by LH47's O stars to the east of Shell 2.  The structure 
of the ionized gas in Shell 2 thus depends of the relative locations 
of the ionizing stars, and would not have a one-to-one correspondence 
with the neutral \hi\ gas shell.

The channel maps of \hi\ in N44 also show numerous holes.  The most
obvious \hi\ holes are present in the map centered at the bulk 
velocity of N44, \vhel\ = 297 \kms.  In addition to the aforementioned
superbubbles Shell 1 and Shell 2, two other holes are present to the
north.  As described in \S 3, one of these \hi\ holes has \ha\ 
filaments delineating a shell structure, while the other has no 
evidence of a coherent shell structure in \ha\ images.  These two
\hi\ holes must be real holes, as both regions show enhanced X-ray
emission (see Figure 5), indicating that the holes are filled with 
10$^6$ K hot gas.  It might be argued that the enhanced X-ray 
emission is resultant from a smaller foreground absorption.  Since 
the integrated \hi\ map does not show obvious anti-correlation 
between \hi\ column density and X-ray surface brightness in N44, 
we consider this argument unlikely.  These apparent \hi\ holes in 
the channel maps must correspond to real cavities.

Conversely, we next examine the \hi\ maps for holes at regions where
we expect them.  The SNR 0523$-$679, having a diameter of $\sim4'$,
is well resolved by our \hi\ maps.  However, only a very shallow 
depression in \hi\ column density is present in the integrated \hi\
map or the channel map centered at 297 \kms.  This is perhaps
understandable because the SNR 0523$-$679 is in a low density medium,
indicated by its fast expansion and large size.  The \hi\ gas 
that has been accelerated by the SNR also shows a patchy structure,
consistent with a low density medium around the SNR.

We may also expect \hii\ regions to show up as holes in \hi\ maps.
However, neither the \hii\ region around LH48 nor the \hii\ region
around LH49 shows a hole in the integrated or channel maps of \hi.
The lack of correspondence between young \hii\ regions and \hi\
distribution indicates that the star formation must have taken place 
in molecular clouds and the \hii\ regions contain the dissociated 
and ionized molecular gas.  This conclusion has been previous 
reached by Allen, Atherton, \& Tilanus (1986) based on their \hi\ 
observations of the spiral galaxy M83.  The presence of molecular 
cloud in N44 is evidenced in the CO map by Cohen et al. (1988).
CO maps made with the ESO-SEST at a higher resolution show molecular
material associated with the B, C, and D components of N44, the 
compact \hii\ regions along the west (the B component) and 
southwest (the C component) rim of Shell 1 and to the southeast 
(the D component, associated with LH 49) of Shell 1 
(\markcite{Is93}Israel et al. 1993; \markcite{Chin97}Chin et al. 
1997).  This molecular material must be the remnants
of the natal clouds. 


Besides \hi\ holes and expanding shells, we see clear evidence of
acceleration of \hi\ gas by superbubbles or SNRs, compression of
\hi\ gas between expanding shells, and breakout structures in 
N44.  Similar phenomena must exist in other active star forming 
regions.  It is conceivable that the interstellar medium has been
shaped by multiple episodes of massive star formation, and that the
complex \hi\ structure represents the cumulative effects of the 
previous massive star formation.  

\subsection{\hi\ Shell and Superbubble Dynamics}

The most remarkable lesson we have learned from the \hi\ study 
of superbubble in N44 is that the neutral gas actually contains 
more mass and kinetic energy than the ionized gas!  For example,
in Shell 1 of N44, the mass and kinetic energy in the expanding 
\hi\ shell is 3--7 times those of the ionized gas shell.  Thus, 
the \hi\ gas is an important component in superbubbles.  
Previously, superbubble dynamics has been observed with optical
emission lines, hence only the ionized gas shells have been 
considered.  This clearly needs to be improved. 

It has been a standing problem that the observed kinetic energy in
a wind-blown bubble is much too low compared to that expected from
Weaver et al.'s (1977) pressure-driven bubble model (Oey \& Massey
1995).  We have summed the kinetic energy in the ionized gas
shell and in the \hi\ shell for the superbubble Shell 1 in N44, 
and found that the observed total kinetic energy is still at least 
a factor of 5 lower than expected.  The observed ratio of kinetic 
energy to thermal energy is nevertheless closer to the theoretical 
value of 3/7, especially if the thermal energy in the breakout 
region is included.  

The superbubble dynamics of Shell 1 may have been significantly
altered by the energy loss through the breakout; however, the 
breakout may not be the only cause for the discrepancy between
the observed superbubble dynamics and theoretical predictions.
The fact that the age of the OB association LH47, $\ge$ 10 Myr
(Oey \& Massey 1995), is much larger than the dynamic age of
Shell 1, $<$ 1 Myr, indicates that the formation of a superbubble 
did not start as soon as the central OB association was formed.
It is possible that LH47 was formed in a molecular cloud, and 
the currently visible superbubble Shell 1 did not start to expand
rapidly until it has broken out of the molecular cloud.  This 
hypothesis is supported by the remnant molecular clouds observed 
toward N44 (Israel et al. 1993; Chin et al. 1997).  Only the 
youngest \hii\ regions are still associated with molecular 
material, and these youngest \hii\ regions show neither \hi\
holes nor \hii\ shell structures.
Similar explanation has been proposed for the superbubble in 
N11, and may be common among all superbubbles (\markcite{Ma97}Mac 
Low et al. 1997). 

Of course, off-center SNR shocks could have provided acceleration
and contributes to the discrepancy between the observed superbubble
dynamics and the theoretical predictions.  Evidence of SNR shocks
in N44 includes the excess X-ray emission (Chu et al. 1993) and
strong [S II]/H$\alpha$ ratios (\markcite{Hu94}Hunter 1994).
Future observational studies of superbubble dynamics need to 
include both the neutral and ionized gas components and extend
to superbubbles with a wide range of X-ray luminosities, so that 
additional SNR acceleration effects can be differentiated from
normal superbubble dynamics.

\begin{acknowledgments}

We thank the team members of the HI survey project -- 
R. Sault, K. C. Freeman, M. A. Dopita, M. J. Kesteven, D. McConnell, 
and M. Bessell. SK would like to thank P. R. Wood for helpful 
comments to improve the paper, and J. Mould and MSSSO for financial 
supports to make the collaborative visit to Illinois possible.  
YHC acknowledges the NASA grants NAG 5-3246 (LTSA) and NAG 5-1900
(ROSAT). RCS acknowledges NSF grant AST-9530747 and the Dean B.
McLaughlin Fund at the University of Michigan.

\end{acknowledgments}

\newpage

\centerline{\bf FIGURE CAPTIONS}

\begin{figure}[th]

\caption{H$\alpha$ image of the N44 complex and echellograms. The 
echelle slit positions are marked in the \ha\ image. The slit `S3' 
samples the center of the superbubble Shell 1.  The slit ``S1'' 
samples the breakout region, or the X-ray South Bar (Magnier et al. 
1996). `S4' and `S5' sample the lobes of the SNR 0523$-$679.
The two narrow lines in the echellograms are the geocoronal H$\alpha$
and the telluric OH line blend at 6577.284 \AA.}

\caption{An ATCA integrated HI map of the N44 complex. The optically 
defined `Shell 1' and `Shell 2' are labeled on the HI contours. 
Corresponding HI holes in these positions can be clearly seen.}

\caption{The ATCA channel maps of N44's HI datacube. The heliocentric 
velocity is given in the upper left corner.  The \hi\ contours are
overplotted on (a) greay-scale images of \hi\ themselves, and (b)
an \ha\ image.  The contour levels are 0.07, 0.10, 0.15, 0.20, 0.25,
0.30, 0.35, 0.40, 0.45 Jy/Beam.}

\caption{The position-velocity plots of HI in the N44 complex. The 
contours corresponds to the associated pixel brightness $\geq$ 2.5 
$\sigma$. The high-velocity gas associated with SNR might have 
unresolved gas component, which relate with an H$\alpha$ blister on 
the periphery of `Shell 1'.}

\caption{\hi\ contours overplotted on an X-ray image of N44.  The
\hi\ contours are derived from the channel map centered at \vhel\ = 
297 \kms.  The \hi\ contour levels are 0.07, 0.10, 0.15, 0.20, 0.25,
0.30, 0.35, 0.40, 0.45 Jy/Beam. The X-ray image is a smoothed ROSAT 
PSPC image from Chu et al. (1993). }

\end{figure}

\clearpage

\begin{table}[h]
\caption[junk]{Journal of Echelle Observations}
\renewcommand{\footnoterule}{}
\begin{minipage}{5.75in}

\begin{tabular}{lllccl}
\tableline \tableline 
No. & RA (J2000) & Dec (J2000)   &  Slit  & Exposure & Remarks \\
    & (hh mm ss) & (~~$^\circ$ ~~$'$ ~~$''$~) & Orientation & (s) &   \\ 
\tableline 
1   & ~05 22 32  &  $-$68 00 06  &  E--W  & 1200  & South breakout region \\
2   & ~05 22 32  &  $-$68 00 46  &  E--W  & 1200  & 40$''$S of slit 1 \\ 
3   & ~05 22 17  &  $-$67 56 18  &  E--W  & 2$\times$1200 & Center of Shell 1 \\
4   & ~05 23 01  &  $-$67 52 02  &  E--W  & 2$\times$1200 & SNR \\ 
5   & ~05 23 05  &  $-$67 54 09  &  N--S  & 2$\times$1200 & SNR \\ 
6   & ~05 22 32  &  $-$68 00 26  &  E--W  & 1200  & 20$''$S of slit 1 \\
\tableline
\end{tabular}
\\
\\
\end{minipage}
\end{table}
\clearpage

\begin{table}[h]
\caption[junk]{The Physical Parameters of Shell 1 and Shell 2}
\renewcommand{\footnoterule}{}

\begin{tabular}{lcccccc}
\tableline \tableline
Parameter               & \multicolumn{2}{c}{Shell 1$^a$} & 
\multicolumn{2}{c}{Shell 1$^b$} & \multicolumn{2}{c}{Shell 2} \\
                             & Front & Back & Front & Back & Front & Back \\
\tableline
The \hii\ hemisphere: & & & & & & \\
Shell Thickness (pc)         &  0.5 &  0.5 &  2.3   & 2.3  &  ---  & --- \\
$n_e$ (cm$^{-3}$)            & 11.9 & 20.7 &  5.6   & 9.7  &  ---  & --- \\
Mass ($M_\odot$)             &  440 & 770  &  950   &1630  &  ---  & --- \\
$V_{exp}$ (\kms)             &   45 &  30  &   45   &  30  &  ---  & --- \\
$E_{kin}$ (10$^{49}$ ergs)   & 0.89 & 0.68 &  1.9   & 1.5  &  ---  & --- \\
\tableline
The \hi\ hemisphere: & & & & & & \\
$N(HI)$ (10$^{20}$ cm$^{-2}$)& 3.6  & 1.4     &    &       &  1.6  &  3.0  \\
Mass ($M_\odot$)             & 6180 & 2460    &    &       &  3090 & 5850  \\
$V_{exp}$ (\kms)             & 28   & 50 - 60 &    &       & 50$\pm$5 & 30$\pm$5 \\
$E_{kin}$ (10$^{49}$ ergs)   & 4.8  & 6.1     &    &       &  7.7  &  5.2  \\
\tableline
The whole shell: & & & & & & \\
\hii\ $E_{kin}$ (10$^{49}$ ergs)    & \multicolumn{2}{c}{1.6 -- 3.4} & & &    & \\
\hi\  $E_{kin}$ (10$^{49}$ ergs)    & \multicolumn{2}{c}{11}         & & 
&\multicolumn{2}{c}{13} \\
Total $E_{kin}$, observed (10$^{49}$ ergs) & \multicolumn{2}{c}{13 -- 14}& & & & \\
$E_{th}$, observed in shell (10$^{49}$ ergs)    & \multicolumn{2}{c}{3 -- 20}     & & & & \\
$E_{th}$, observed in shell \& breakout (10$^{49}$ ergs)  & \multicolumn{2}{c}{13 -- 30}     & & & & \\
Wind and SN Energy Input (10$^{49}$ ergs)& \multicolumn{2}{c}{400 -- 700}& & & & \\
Total $E_{kin}$, expected (10$^{49}$ ergs)&\multicolumn{2}{c}{78 -- 136} & & & & \\
$E_{th}$, expected (10$^{49}$ ergs)    & \multicolumn{2}{c}{200 -- 300}  & & & & \\
\tableline
\end{tabular}

\medskip
\medskip
\noindent Notes: \\  
\noindent $^a$ with an ionized shell thickness $\Delta R/R$ of 0.02 \\
\noindent $^b$ with an ionized shell thickness $\Delta R/R$ of 0.1
\end{table}

\end{document}